\documentclass{jpp}
\usepackage{graphicx}

\usepackage[utf8]{inputenc}
\usepackage[T1]{fontenc}
\usepackage{amsmath}

\shorttitle{Numerical simulations of relativistic jets}
\shortauthor{Manel Perucho and J. López-Miralles}

\title{Numerical simulations of relativistic jets}

\author{Manel Perucho\aff{1,2,3}
  \corresp{\email{manel.perucho@valencia.edu}} and
  J. López-Miralles\aff{1,4}
  }

\affiliation{\aff{1}Departament d'Astronomia i Astrof\'{\i}sica, Universitat de Val\`encia, C/ Dr. Moliner, 50, 46100, Burjassot, Valencian Country, Spain.
\aff{2}Observatori Astron\`omic, Universitat de Val\`encia, C/ Catedr\`atic Jos\'e Beltr\'an 2, 46980, Paterna, Valencian Country, Spain.
\aff{3}Institut für Theoretische Physik und Astrophysik, Universität W\"urzburg, Emil-Fischer-Straße 31, 97074 W\"urzburg, Germany.
\aff{4}Aurora Technology for the European Space Agency, XMM-Newton Science Operations Center, ESAC/ESA, Camino Bajo del Castillo s/n, Urb. Villafranca del Castillo, 28691 Villanueva de la Cañada, Madrid, Spain}

\begin{document}

\maketitle

\begin{abstract}
In this paper, we review recent and ongoing work by our group on numerical simulations of relativistic jets. Relativistic outflows in Astrophysics are related to dilute, high energy plasmas, with physical conditions out of the reach of current laboratory capabilities. Simulations are thus imperative for the study of these objects. We present a number of such scenarios that have been studied by our group at the Universitat de Val\`encia. In particular, we have focused on the evolution of extragalactic outflows through galactic and intergalactic environments, deceleration by interaction with stars or clouds, or the propagation of jets in X-ray binaries and interaction with stellar winds from massive companions. All also share their role as particle acceleration sites and production of non-thermal radiation throughout the electromagnetic spectrum. Therefore, our work is not only aimed to understand the impact of outflows on their environments and thus their role in galaxy and cluster evolution, but also the nature and capabilities of these sites as generators of high and very-high energy radiation and cosmic rays. 
\end{abstract}

\section{Introduction}

Jets and outflows represent a common and crucial process in Astrophysical systems; from forming stars (Herbig-Haro objects) to supermassive black holes in active galactic nuclei (AGN). In each of these systems, the properties of the jets (velocities, densities, composition...) are different, but they share a common association to accretion and magnetohydrodynamical processes that explain their formation, collimation and acceleration \citep[see, e.g.,][]{2012rjag.book...81K}. 

We know that extragalactic, AGN jets, can be modelled as fluids or plasmas because the typical Larmor radii of the jet particles is orders of magnitude smaller than the spatial scales of the system \citep{1974MNRAS.169..395B}. The same is true for the case of jets generated in collapsars and X-ray binary stars. Despite technical advances, it is still impossible to reproduce the most extreme of these systems, i.e, accelerate plasmas to relativistic velocities, in laboratories to study their properties \citep[e.g.,][]{2018JPlPh..84e7501B}.\footnote{This would imply including enough particles to provide the system with fluid/plasma properties.} This leaves numerical simulations as the only option to approach the nature of these systems, which can be central in the evolution of the interstellar medium in host galaxies and clusters \citep[e.g.,][]{2007ARA&A..45..117M,2012ARA&A..50..455F}, or in the generation of high and very-high energy radiation in the Universe \citep[e.g.,][]{2018Galax...6..116R}. 

Relativistic jets are formed in the environment of compact objects by the extraction of rotational energy from the accreting body and/or the accretion disk \citep{1977MNRAS.179..433B,1982MNRAS.199..883B}. The idea is, in both cases, related to the twisting of magnetic field lines, which by resisting to this deformation, are able to extract rotational energy from the system. The magnetic field is supposed to be dragged onto the central object by the accreting gas itself. In the case of Kerr black holes, it anchors to the hole's ergosphere and is thus forced to rotate, faster inside than outside of this region, which creates the aforementioned twisting \citep{1977MNRAS.179..433B,2012rjag.book...81K}. 

In all cases, jet evolution involves many orders of magnitude in distance. In the case of AGN jets, jet formation takes place in spatial scales of the order of the gravitational radius of the central supermassive black hole (SMBH), i.e., $\sim 10^{10}\,{\rm m}$, or, if the inner regions of the accretion disk are involved, it may extend to a factor $~10-100$ larger. They reach, in contrast, distances up to $\sim 1~\,{\rm Mpc}, \mathrm{i.e.} \sim 10^{22}\,{\rm m}$, as revealed by low frequency surveys \citep[see, e.g.,][]{1991bja..book..232I,2014A&A...565A...2M,2022A&A...660A...2O,2023A&A...672A.178S}. As a consequence, the computational resources to study the whole evolution are prohibitive even for modern architectures and the studies focus on different regions. Depending on this, the properties of the jets vary and the numerical codes have to be adapted to them.

When ejected, jets consist mainly of a Poynting flux, which may be loaded by a dilute, hot gas with an electron-positron pair composition –the pairs would be generated by photon-photon collisions at the black hole corona, and/or an electron-proton gas if loaded from the accretion disk. These jets are launched with a relatively small radial velocity. Taking into account the large values achieved by magnetic fields in these regions, their velocities must initially be sub-Alfvènic. Nevertheless, relativistic speeds are reported at sub-parsec to parsec scales, which means that jets need to be accelerated to super-Alfvènic (super-fast-magnetosonic) speeds. The mechanisms that are typically invoked to explain this acceleration are both related to jet expansion with magnetic and internal energy contributions \citep[e.g.,][and Ricci et al., in preparation]{2004ApJ...605..656V,2007MNRAS.380...51K}.

At kiloparsec scales, jets show different morphologies and clear signs of interaction with the warm-hot intergalactic medium \citep[WHIM, e.g.,][]{2019A&A...622A..10C}. A coarse, initial dichotomy was established by \cite{1974MNRAS.167P..31F} between Fanaroff-Riley type I (FRI) sources and Fanaroff-Riley type II (FRII) sources. The former are brighter within their host galaxies, jets become dimmer with distance and show brightness symmetry at large distances \citep[e.g.,][]{2014MNRAS.437.3405L}. FRIIs are brighter at the impact sites with the WHIM, the so called hotspots, and show brightness asymmetry along their length between the jets and their counter-jets. The common interpretation is related to jet velocity: on the one hand, FRII jets seem to keep their collimation and mildly relativistic velocities up to the interaction site with the WHIM, which means that the flow is supersonic at impact, triggering strong shocks that show up as hotspots, where Doppler boosting explains the brightness asymmetry; on the other hand, in the case of FRI jets, the loss of collimation and brightness symmetry at large scales point towards deceleration and weakening of the Doppler asymmetry \citep{1984AJ.....89..979B,1984ARA&A..22..319B}.

In the case of microquasars, although jets evolve in time-scales much shorter than AGN jets \citep[see, e.g.,][for a review]{mirabel, Fender}, a large number of unknowns remain about fundamental aspects such as their composition \citep{migliari02}, internal dynamics and its relation to the forming mechanism and changes in the accretion flow \citep{fender04}, or the mechanisms that generate the production of high to very-high energy radiative output within the binary scales \citep[see, e.g.,][]{2008A&A...482..917P,2009IJMPD..18..347B,2010A&A...512L...4P,2012A&A...544A..59B,2012A&A...539A..57P,2015A&A...577A..89B,2015A&A...574A..77P,2017A&A...598A..13D,2022A&A...661A.117L}, even far from the sub-miliarscsecond resolution achieved by very large baseline interferometry (VLBI) at cm-mm wavelengrths in the radio band. Furthermore, the transition of the jet from these compact regions ($\sim 10^{10}\,{\rm m}$) to parsec scales ($\sim 10^{16}\,{\rm m}$) has been barely studied by numerical simulations, and only few works have been devoted to microquasar jet evolution at the largest scales \citep[e.g.,][]{2009A&A...497..325B,monceau,charlet}. 

In all scenarios in which relativistic outflows are involved, and once accelerated, the jets are subject to dissipative processes such as shocks or mass entrainment via interaction with stars and development of instabilities \citep{2019Galax...7...70P}. The remarkable stability of jets along the different scales through which many of them are observed to propagate has been a matter of study during the last decades. The mechanisms by which extragalactic, FRI jets are decelerated have also been deeply investigated \citep[e.g.,][]{2014MNRAS.441.1488P, 2016A&A...596A..12M,2019A&A...621A.132M,2022A&A...659A.139M,2018NatAs...2..167G,2020MNRAS.494L..22P,2021MNRAS.500.1512A}. Attention has been paid to the scenarios of generation of high energy particles \citep[e.g., neutrinos, see][]{2022arXiv220203381M,2023arXiv230511263B} and cosmic rays in these systems, both in galactic and extragalactic sources \citep[e.g.,][and references therein]{2018Galax...6..116R,2020NewAR..8901543M,2022arXiv221204159S}. The particles could be accelerated via shock or shear acceleration, or also magnetic reconnection processes, i.e., as a byproduct of jet development. Recent works based on particle-in-cell simulations suggest a relation between synchrotron light polarization and jet composition \citep[e.g.,][]{2021Univ....7..450M}. It is relevant to stress that a baryonic component is necessary to explain the generation of neutrinos, for instance. This component could be either provided by a disk-generated wind surrounding the inner leptonic jet spine, or by entrainment. Finally, the role of jets in heating galactic environments, avoiding the fast cooling and collapse of the WHIM onto the jet's host galaxy, also known as feedback process, could be very relevant in galaxy evolution and is not only part of research on jet physics, but also in cosmological simulations \citep[e.g.,][]{2017MNRAS.470.1121T,2017MNRAS.470.4530W,2023A&A...669A..50V}. 

In this paper, we summarize our recent work on jet physics, which is mainly focused on jet propagation, stability, interaction with the ambient medium at different scales in both extragalactic sources and X-ray binaries, via numerical simulations, but also present new results either based on revisiting published work or on new simulations. In section 2, we present the basic equations that need to be solved to simulate the evolution and physics of these systems, and the numerical techniques that we use to solve them. Section 3 is devoted to the different applications that we have performed in the last years, both in the case of relativistic hydrodynamics (RHD) and magnetohydrodynamics (RMHD), and the main results that we have obtained. Among these, we include new results such as the pseudo-synchrotron images derived from AGN jet simulations (Figs.~\ref{j300} and \ref{j310} in Sect.~\ref{sec:frii}), or the new simulations on jet-interstellar medium (ISM) interactions (Sect.~\ref{sec:ism}), and new analysis of recent RMHD simulations (Sect.~\ref{sec:rmhd}). Finally, our conclusions are given in section 4. 

\section{Numerical codes and methods}

\subsection{RMHD equations}
The stress-energy tensor for an ideal plasma is $T^{\mu \nu} = T^{\mu \nu}_G + T^{\mu \nu}_{EM}$. On the one hand, the gas contribution is given by:
  
\begin{equation}
T^{\mu \nu}_G = \rho h u^{\mu} u^{\nu} + p g^{\mu \nu},
\end{equation}
where the Greek letters take values $\mu,\,\nu=0,1,2,3$ and stand for the dimensions of the Minkowski space-time, $\rho$ is the flow rest-mass density, $h$ the gas specific enthalpy, $u$ is the four-velocity, $p$ is the gas pressure, and $g$ the Minkowski metric, and we have used $c=1$. On the other hand, the field contribution is given by:

\begin{equation}
T^{\mu \nu}_{EM} = F^\mu_\beta F^{\nu \beta} \,-\, \frac{1} {4} g^{\mu \nu} F_{\alpha \beta} F^{\alpha \beta},
\end{equation}
where $F^{\alpha \beta}=\epsilon^{\alpha\beta\mu\nu} b_\mu u_\nu$ are the elements of the electromagnetic field tensor satisfying the Maxwell equations:
\begin{equation}\label{max}
\partial_{\alpha} F_{\mu \nu} = 0, \qquad \nabla_\mu F^{\mu \nu} = -J^\nu.
\end{equation}
\noindent
In this expression $\epsilon^{\alpha\beta\mu\nu}$ is the Levi-Civita tensor, $b_\mu$ are the components of the magnetic field in the fluid rest-frame, and $J^\nu$ is the charge four-current. Collecting the expressions above, we obtain the stress-energy tensor of a magnetized, perfect fluid:
\begin{equation}
T^{\mu \nu} = \rho h^* u^\mu u^\nu + p^* g^{\mu \nu} - b^\mu b^\nu,
\end{equation}
where $h^*= h + |b|^2 /\rho$ is the total enthalpy, $p^*$ the total (gas plus magnetic) pressure ($p^*=p_{\rm gas} + |b|^2/2$), $b^\mu$ is the relativistic magnetic field four-vector, with $b^0=\gamma(\boldsymbol{v}\cdot\boldsymbol{B})$, and $b^i=\frac{B^i}{W}+v^ib^0$, so $|b|^2 = b_\alpha b^\alpha= B^2/\gamma^2 + (\mathbf{v}\cdot\mathbf{B})^2$, with $\mathbf{B}$, $\mathbf{v}$ the magnetic field and velocity vectors in the laboratory frame, and $\gamma$ the corresponding Lorentz factor. We note that a $\sqrt{4\pi}$ factor is included into the definition of the magnetic fields.

The conservation laws that govern the dynamics of the plasma are:

\begin{equation}
\nabla_\mu\,T^{\mu \nu} = 0, \qquad  \nabla_\mu (\rho u^\mu) = 0.
\end{equation}  

These equations can be written in matrix form as:

\begin{equation}
    \partial_t\boldsymbol{U}+\partial_i\boldsymbol{F}^i=0,
    \label{eq0}
\end{equation}
where $\boldsymbol{U}=\{D,S^j,\tau_e,B^j\}$ is a vector of conserved variables, $D$ is the relativistic rest mass density, $S^j$ is the momentum density of the magnetized fluid, $\tau_e$ is the energy density, 
and $\boldsymbol{F^i}$ are the fluxes in each spatial direction. The relation of those conserved quantities and fluxes to their primitive physical variables is given by:

\begin{equation}
\label{eqU}
\boldsymbol{U}=
    \begin{pmatrix}
D\\ 
S^j\\ 
\tau_e\\ 
B^j\\ 

\end{pmatrix}=
    \begin{pmatrix}
\rho \gamma \\ 
\rho h^* \gamma^2v^j-b^0b^j\\ 
\rho h^* \gamma^2-p^*-b^0b^0-\rho \gamma\\ 
B^j\\ 
\end{pmatrix},
\end{equation}
\begin{equation}
\label{eqF}
\boldsymbol{F^i}=
        \begin{pmatrix}
\rho \gamma v^i \\ 
\rho h^* \gamma^2v^iv^j+p^*\delta^{ij}-b^ib^j\\ 
\rho h^* \gamma^2v^i-b^0b^i-\rho \gamma v^i\\ 
v^iB^j-B^iv^j\\ 
\end{pmatrix}.
\end{equation}

Finally, from Eq.~\ref{max}, we obtain the divergence-free condition for the magnetic field, $\nabla \cdot \boldsymbol{B} = 0$. 

This system of equations needs to be closed by means of an equation of state. It is common to use the ideal gas equation of state for this purpose. However, in the case of relativistic outflows propagating through their host galaxies, the different composition of the jet flow (electron/positron pairs and/or electron/proton) and the ambient medium (ionized hydrogen in the intergalactic medium, but also atomic hydrogen in two phase media such as the interstellar medium) recommend the use of the relativistic gas equation of state \citep{synge1957}. 

Source terms can be added to the equations, depending on the problems to be solved. Typical source terms used are gravity \citep[or pseudo-gravity to keep a density/pressure profile in equilibrium, e.g.][]{2007MNRAS.382..526P,2019MNRAS.482.3718P}, thermal cooling terms \citep[e.g.,][]{2017A&A...606A..40P}, hydrogen ionization and recombination effects \citep[e.g.][Perucho et al., in preparation]{2015A&A...580A.110V,2021AN....342.1171P}, or radiation feedback \citep{2023CoPhC.28408630L}.

It is worth noting that, on the one hand, relativistic effects arise when $v\rightarrow c$ and/or when $\varepsilon \rightarrow c^2$, and, on the other hand, magnetic fields can be dynamically relevant for even small values of the magnetic parameter $\sigma=|b^2|/(\rho\,h)$, which gives the ratio of magnetic to gas kinetic energy.

\subsection{The numerical codes}

Numerical codes are aimed to solve the RMHD equations \citep[see][for reviews]{martiRev,2019Galax...7...24M}. In our case, we use multidimensional High Resolution Shock-Capturing methods, with a finite volume scheme. Time advance of the equations is done via dimensional splitting to compute the numerical fluxes across cell boundaries, using the integral form of the conservation equations \citep[see e.g.,][]{komi99}, and Runge-Kutta methods \citep{shu89}.

In the case of RHD simulations, we use our code \textsc{ratpenat}, a hybrid parallel code  -- MPI + OpenMP --\citep[see][and references therein]{2010A&A...519A..41P} in which: i) primitive variables within numerical cells are reconstructed using PPM routines, ii) numerical fluxes across cell interfaces are computed with Marquina flux formula, iii) advance in time is performed with third order TVD-preserving Runge-Kutta methods. 

RMHD simulations are run with the code \textsc{lóstrego} \citep{2022A&A...661A.117L}, which uses the same basic strategy, although with different schemes: i) primitive variables are reconstructed using piecewise linear methods (PLM): MinMod \citep{roe86}, MC \citep{VanLeer77} or VanLeer \citep{vanleer74}, with slope limiters that preserve monotonicity, or monotonicity-preserving methods \citep[MP;][]{suresh97}, (ii) the approximate solvers used to compute the fluxes are HLL \citep{harten83}, HLLC \citep{Mignone06} and HLLD \citep{Mignone09}. In this case, the divergence free condition of the field is maintained by using the constrained transport technique \citep[CT,][]{evans88,Ryu98,balsara99,gardiner05}. 

These codes are run in supercomputing resources at the Universitat de Val\`encia (Tirant, Vives) or within the the Spanish Supercomputing Network (Red Espa\~nola de Supercomputaci\'on, RES). The parallelization used in both codes permits a three-dimensional splitting of the grid for distributed memory architectures. In this contribution, we present simulations which typically use 512-1024 cores and require $>\,100\,{\rm khr}$ computing hours.

\section{Applications}
\subsection{RHD simulations}
\subsubsection{Long term evolution} \label{sec:frii}
One line of our work has been aimed to understand the evolution of powerful jets through the density/pressure profile of their host galaxies, and what is the role they play in heating the intergalactic medium. Our simulations were setup as an initial grid covering $\sim 100-200\,{\rm kpc}$ in the three-dimensional case \citep{2019MNRAS.482.3718P,2022MNRAS.510.2084P} to $\sim 1\,{\rm Mpc}$ in two-dimensional simulations \citep{2011ApJ...743...42P,2014MNRAS.441.1488P}. The jets are injected as a boundary condition, with properties expected for jets at the injection point, typically 1~kpc from the central formation engine. The ambient density at this distance is $\sim 10^{5}\,{\rm m^{-3}}$, and it drops with distance following a profile derived from modelling X-ray observations of radio galaxy 3C~31 \citep{2002MNRAS.334..182H}.

At these scales, the jet is expected to be dominated by kinetic energy flux, and the magnetic field to have a dominant disordered component, thus only contributing to pressure, but triggering minor tension forces. This is an assumption that we apply to simulate these scenarios with RHD codes.

The jets are injected in the grid with mildly relativistic velocity $0.9\,c$ and specific enthalpies around $c^2$. All these simulations show that collimated, relativistic jets generate high-pressure regions of shocked jet gas. The high sound speed in this region facilitates equilibrium within the whole shocked region, feeding pressure-driven shock expansion. Altogether, this is a very efficient process in terms of energy transfer from the jet to the ambient medium, as explained in \cite{2017MNRAS.471L.120P}. The main point that the simulations revealed is that because the post-shock pressure depends on the momentum flux density, collimated, relativistic jets transfer most of their energy flux (which is not dominated by the rest-mass energy of the particles, as it happens in classical jets) to the ambient medium via shock heating. 

\begin{figure*}%
    \includegraphics[width=0.5\linewidth]{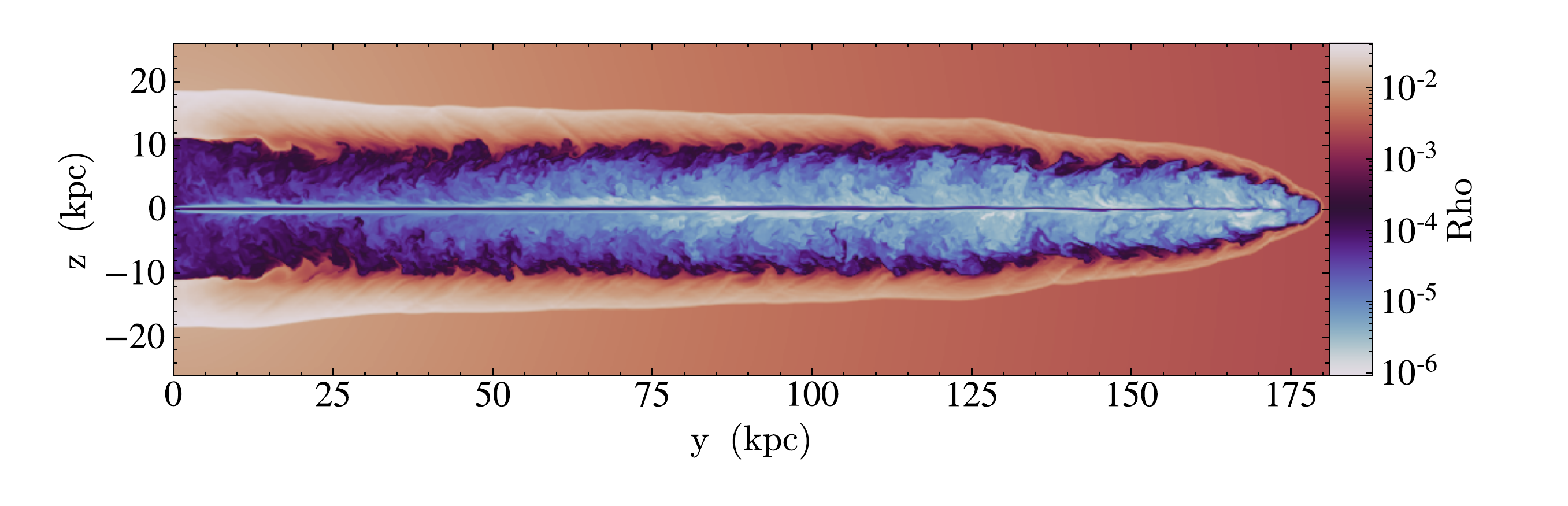}
    \includegraphics[width=0.5\linewidth]{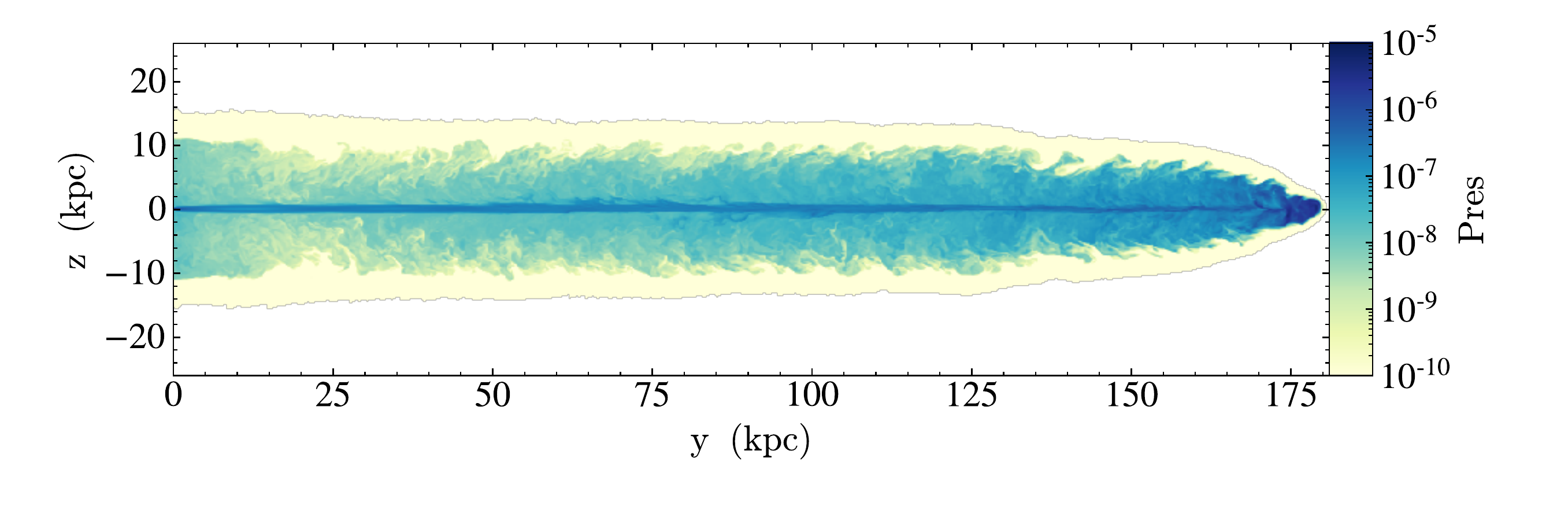} 
    \includegraphics[width=\linewidth]{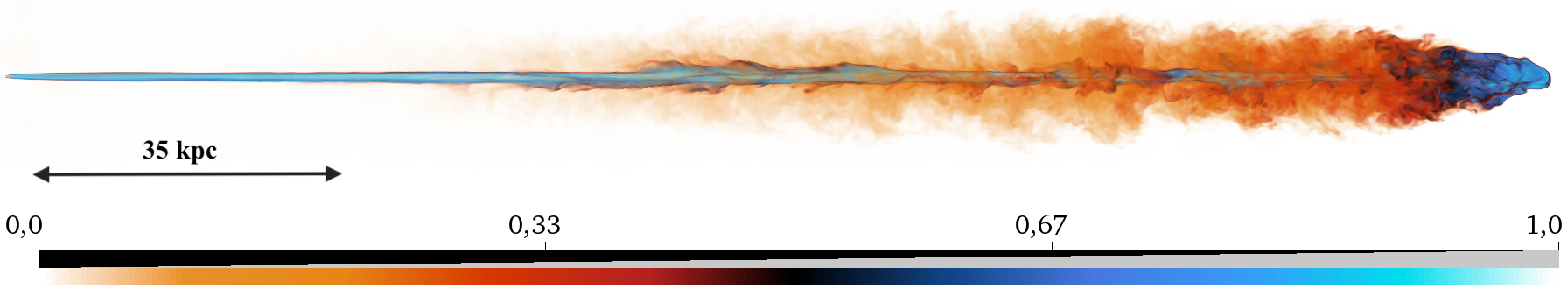} 

    \caption{Top-left panel: Rest-mass density (units: $\rho_a\,=\,1.7\times10^{-22}\,{\rm kg/m^3}$) cut. Top-right panel: Pressure weighted by tracer (units: $\rho_a\,c^2\,=\,1.6\times10^{-5}\,{\rm Pa})$. Bottom: tracer rendering (jet mass-fraction, ranging between 0 for the ambient gas, and 1 for the jet gas).}
  \label{j300}
\end{figure*}

Jet collimation is maintained due to its propagation through a decreasing density/pressure environment, which favours a faster decrease in the pressure of the shocked region, allowing 1) the jet to expand with small opening angles, and 2) the increase of the relative value of jet-to-ambient inertia, which limits the growth rates of unstable modes \citep{1982ApJ...257..509H}. Furthermore, the jet advance velocity through such an ambient medium also increases with distance as compared to a homogeneous, constant density jet. All these factors also contribute to a decrease of the growth rates of unstable modes \citep{2005A&A...443..863P,2019Galax...7...70P}, and to faster expansion, thus allowing the jet to develop to large scales while keeping its collimation.

In summary, our simulations have shown that powerful relativistic outflows ($L_j \geq 10^{38}\,{\rm W}$) propagating through galactic atmospheres are extremely efficient in heating them because of 1) their energy flux is not dominated by rest-mass energy of the particles, and 2) they can keep their collimation.

\begin{figure*}%
    \includegraphics[width=0.5\linewidth]{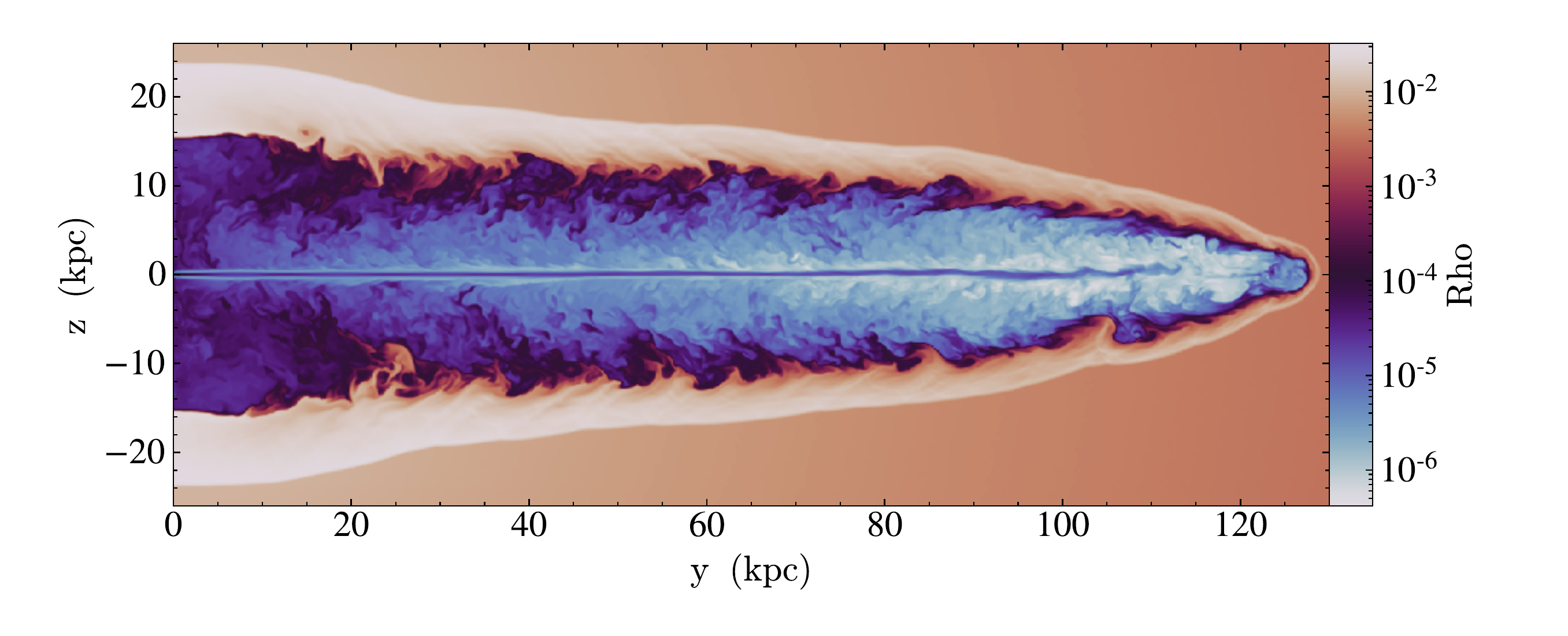}
    \includegraphics[width=0.5\linewidth]{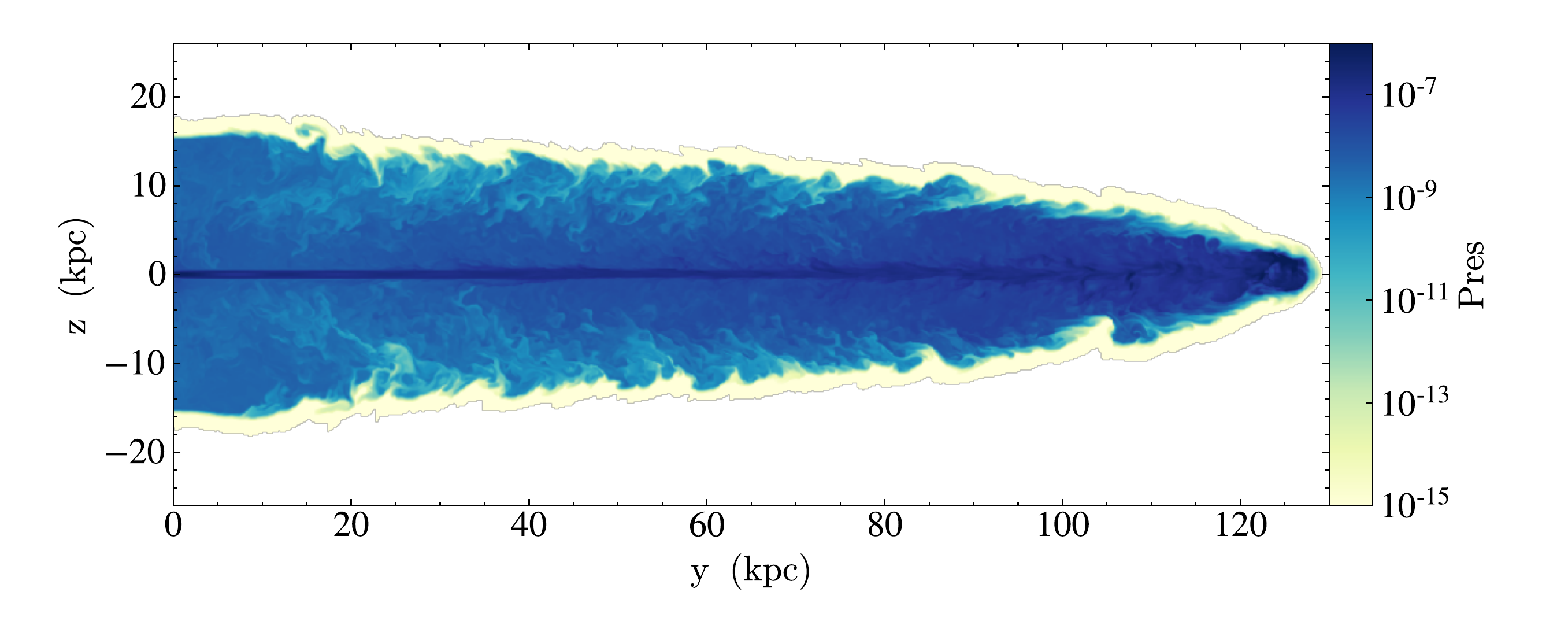} 
    \includegraphics[width=1.0\linewidth]{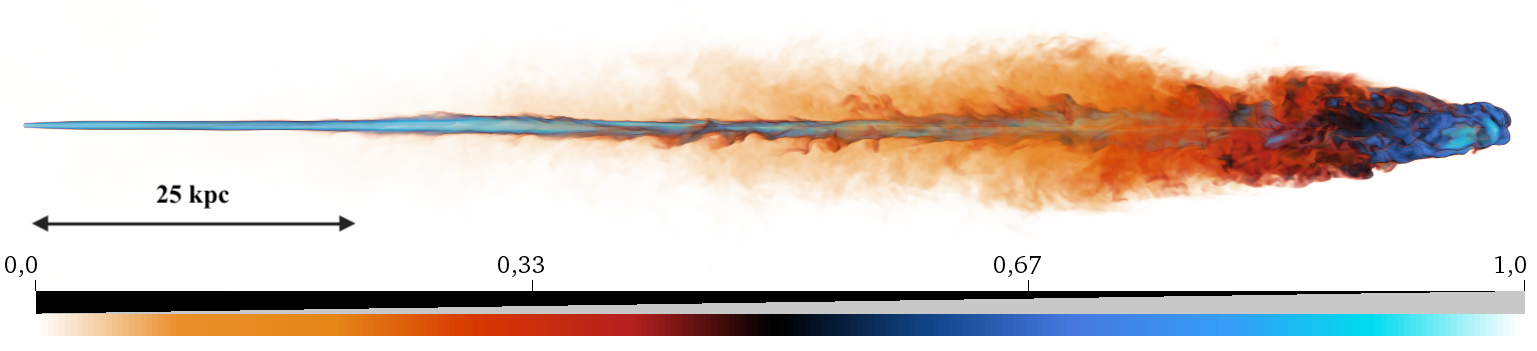} 

    \caption{Top-left panel: Rest-mass density (units: $\rho_a\,=\,6.8\times10^{-22}\,{\rm kg/m^3}$) cut. Top-right panel: Pressure weighted by tracer (units: $\rho_a\,c^2\,=\,6.4\times10^{-5}\,{\rm Pa})$. Bottom: tracer rendering (jet mass-fraction, ranging between 0 for the ambient gas, and 1 for the jet gas). This figure shows the result from a simulation with the same jet properties as those of the jet shown Fig.~\ref{j300}, but with a factor 4 denser ambient medium.}
  \label{j310}
\end{figure*}

Figures~\ref{j300} and \ref{j310} show the results from two of the described simulations, with different ambient medium core density: $10^{-7}\,{\rm m^{-3}}$ in the former and $4\times 10^{-7}\,{\rm m^{-3}}$ in the latter. The images show cuts of rest-mass density and pressure -weighted with tracer- and a rendering of jet-mass fraction (tracer) that shows the jet structure for the last snapshot of each simulation at times $\simeq 5.4\,{\rm Myr}$ and $\simeq 8\,{\rm Myr}$, respectively. In both cases, the jets generate thin backflow/cocoon regions, as caused by the fast propagation through the density decreasing galactic atmosphere. The increased ambient density triggers slightly inflated backflow regions around the jet's terminal shock.

\begin{figure*}%
    \includegraphics[width=\linewidth]{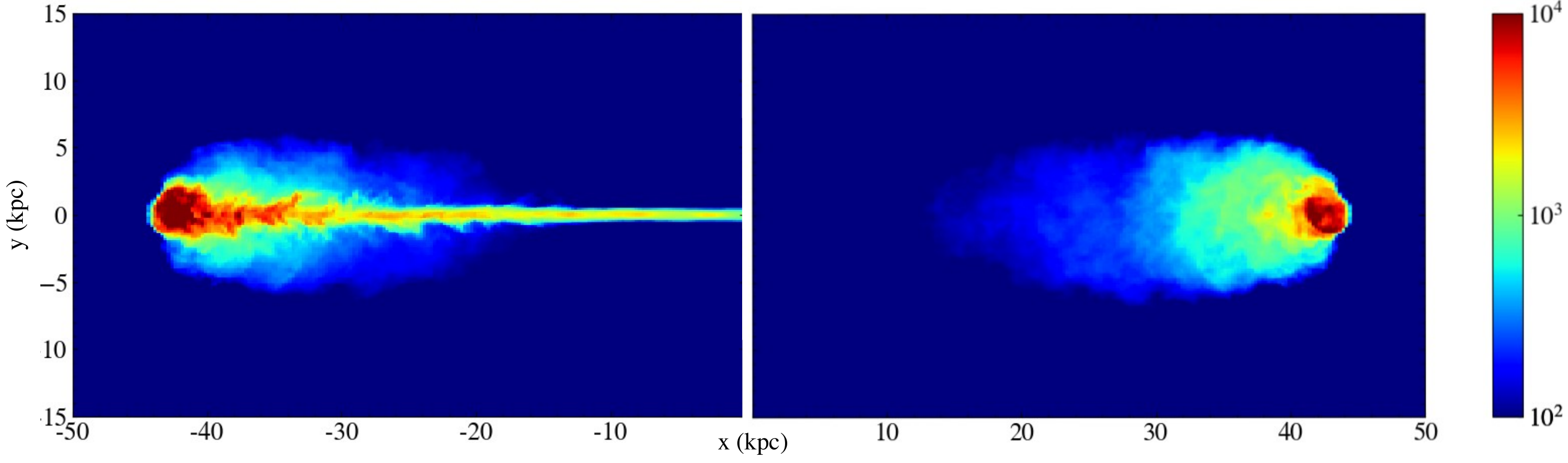}
    \includegraphics[width=\linewidth]{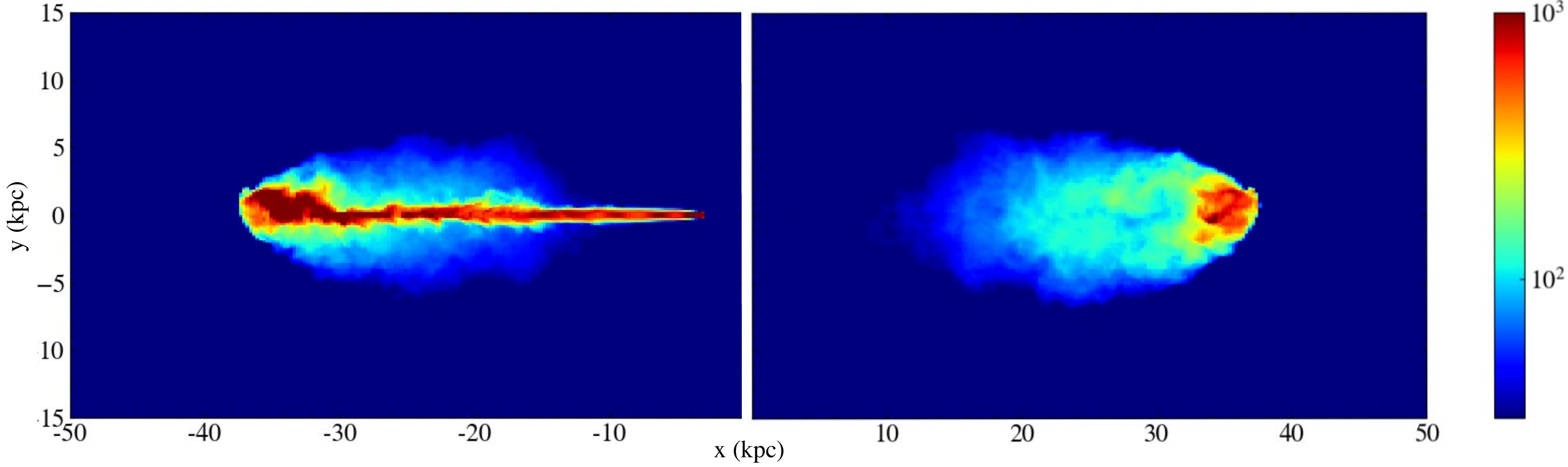} 
    \caption{Pseudosynchrotron emissivity images of the jets shown in Figs.~\ref{j300} (top) and \ref{j310} (bottom) computed with Eq.~\ref{eq:emi}, in arbitrary units.}
  \label{emiss}
\end{figure*}

Figure~\ref{emiss} shows projected pseudosynchrotron emissivity images (at viewing angles of $15^\circ$) of the 3D RHD simulations displayed in Figs.~\ref{j300} and \ref{j310}, from a dynamical point of view in \citet{2019MNRAS.482.3718P,2022MNRAS.510.2084P}. The expression used to compute the emissivity at a given frequency is the same as that used by \citet{2003ApJ...597..798H} \citep[see also][]{1989ApJ...342..700C}: 
\begin{equation}\label{eq:emi}
\epsilon_\nu \,\propto\,n^{1-2\alpha}\,p^{2\alpha}\,(B\,\sin\theta_B)^{1+\alpha}\,D^{2+\alpha},
\end{equation}
where $n$ is the particle number density, $p$ is pressure, $B$ is the assumed magnetic field strength, taken to be proportional to $\sqrt{p}$, $\theta_B$ is the angle between the field lines and the viewing angle assumed to be a constant, $D$ is the Doppler factor, and $\alpha$ is the spectral index (defined as $S_\nu \propto \nu^{-\alpha}$), taken as constant and equal to 0.5 \citep[see, e.g.][]{1988MNRAS.235..997H,1999JPhG...25R.163K}. This expression represents a raw approximation, with strong simplifications as that relative to the magnetic field structure. This assumption cancels differences in emissivity caused by changes in $\theta_B$, but this is expected to be a second order contribution.\footnote{The field is expected to be mainly aligned with the jet in powerful outflows \citep[e.g.,][]{1984AJ.....89..979B}, perpendicular to the jet at the hotspot, randomly oriented in the lobes, and aligned with the bow shock at the outermost surface.} Therefore, the approach is enough to show relevant features. Furthermore, a detailed distribution of the magnetic field across the grid would require a fully RMHD simulation, which is left for future simulations of AGN jets (see Sect.~\ref{sec:rmhd}). The images show the effect of the Doppler factor in generating the jet-to-counterjet brightness asymmetry and reproduce the canonical FRII jet morphology \citep[see, e.g.,][]{1984AJ.....89..979B,1984ARA&A..22..319B,2016MNRAS.458.4443H}, with emissivity enhancements at recollimation shocks, and bright hot-spots surrounded by radio lobes.

\subsubsection{Mass-load}
Less powerful jets ($L_j \leq 10^{36}\,{\rm W}$) can be decelerated by mass-entrainment \citep{1984ApJ...286...68B}. This process may take place in different ways, but mainly through the development of (Kelvin-Helmholtz, current-driven, centrifugal...) instabilities at the jet boundary \citep[e.g.,][]{2007A&A...475..785M,2007MNRAS.382..526P,2013ApJ...772L...1M,2017MNRAS.472.1421M,2018NatAs...2..167G,2016A&A...596A..12M,2019A&A...621A.132M,2022A&A...659A.139M} or by interactions between the jet and stars or clouds that cross it while orbiting the galactic centre \citep[e.g.,][and references therein]{1994MNRAS.269..394K,1996MNRAS.279..899B,2002MNRAS.336.1161L,2006MNRAS.371.1717H,2014MNRAS.441.1488P,2019Galax...7...70P,2021MNRAS.500.1512A}. 

Modelling of FRI jets has revealed that the region in which jets seem to become symmetric in brightness with respect to their counter-jets is also a region in which they become bright and that deceleration takes place progressively, from the jet boundaries towards their axis \citep{2014MNRAS.437.3405L}. This indicates that deceleration is driven by small-scale processes that take place at the boundaries and points towards either small wavelength, instability modes arising at the shear transition with the jet environment or towards other kind of perturbative interactions \citep{2017A&A...606A..40P,2020MNRAS.494L..22P}.

In the case of clouds and stars interacting with the jet, a bow-shock is formed against the jet flow, and the mixing and loading process takes place at the cometary tail formed downstream of the interaction site \citep{1994MNRAS.269..394K}. It has been suggested that this process alone could decelerate low power jets \citep[$L_j \leq 10^{35}\,{\rm W}$,][]{1996MNRAS.279..899B,2014MNRAS.441.1488P} and also change the properties (composition, magnetisation...) in intermediate power jets \citep[$L_j \sim 10^{36}\,{\rm W}$,][]{2021MNRAS.500.1512A}. These changes would be forced, according to \citet{2021MNRAS.500.1512A}, by the relevant relative contribution of the rest-mass density of protons with respect to the jet's injected electron-positron pairs, on the one hand, and the correspondent increase in kinetic energy flux and gas pressure, which reduces the dynamical relevance of the magnetic field, on the other hand.

Furthermore, these interactions can host particle acceleration and generate the production of high-energy radiation and, altogether, may produce the brightness flaring in decelerating jets \citep[e.g.,][]{1997MNRAS.287L...9B,2012A&A...539A..69B,2013A&A...558A..19W,2015MNRAS.447.1001W,2017A&A...604A..57V,2019A&A...623A..91T,2020MNRAS.494L..22P}.

Numerical simulations have allowed us to characterize these scenarios and estimate the radiative output from them at different jet scales, e.g., parsecs and hundreds of parsecs \citep{2012A&A...539A..69B,2017A&A...606A..40P}. Future work should be focused on understanding the role of magnetic fields in these collisions and the study of particle acceleration and cosmic ray production \citep[e.g.,][]{2013A&A...558A..19W,2018ApJ...865..124M}.

\subsubsection{Jet-ISM interactions} \label{sec:ism}
Another aspect that has improved during the last decade in the field of AGN jet simulations is the characterization of the galactic ambient media, both inside and outside the host galaxies \citep[see, e.g.][]{2012ApJ...757..136W,2016MNRAS.461..967M,2018MNRAS.475.3493B,2021AN....342.1171P,2021MNRAS.508.4738M}. Galactic environments have started to be set up by using distributions obtained from cosmological simulations \citep[e.g.,][]{2006MNRAS.373L..65H,2021MNRAS.508.5239Y}. Within host galaxies, the two phase nature of the ISM, with cold, denser gas hosted in clouds and hot, more dilute gas between them represents a non-negligible property to consider. This is specially relevant within the inner kiloparsec of host galaxies, where the broad and narrow line regions are located. 

In addition, observational results at radio-to-UV ranges show that the jets trigger line emission and gas motions in their host galaxies \citep{2018A&ARv..26....4M}. Basing on previous work by \citep{2015A&A...580A.110V}, we have included hydrogen ionization and recombination feedback effects in our RHD code with the aim to trace the effect of jets on the cold gas in the inner kiloparsec \citep{2021AN....342.1171P}. We have now improved the set up, using realistic ambient/cloud densities and temperatures. In these simulations, a purely leptonic ($e^-/e^+$) jet is injected into an inhomogeneous medium composed of a mixture of clouds that host atomic hydrogen, at temperatures $\sim 100\,{\rm K}$ and maximum density $n\sim 10^9\, m^{-3}$, and an ionized medium with $T\sim 10^6\,{\rm K}$ and $n\sim 10^{5}\, m^{-3}$. The numerical box reproduces the inner 500\,pc in the host galaxy, with the jet injected as a boundary condition at a certain distance from the forming region with relativistic velocity $0.98\,c$, density $\rho_j=1.67\times10^{-26}\,{\rm kg/m^3}$, and a very high specific internal energy $\varepsilon\sim10^3\,c^2$ to simulate the high pressure expected in this region (the simulation does not include a magnetic field so this approach could only account for a magnetic pressure generated by a disordered field configuration). In the simulations, the ambient medium is also given a transversal velocity of $100$~km/s to emulate the rotation around the galactic nucleus. 

Altogether, the parameter range in the simulations spans throughout more than eight orders of magnitude in density and pressure, which makes the simulation extremely challenging. Moreover, time-step limiters need to be included to allow for the short hydrogen recombination rates in cold regions or ionization in shocked cells. The simulation uses two equations of state: relativistic \citep{synge1957} and non-relativistic \citep[see][]{2015A&A...580A.110V} ideal gas. The criterion to choose between them in each cell is based on the composition, namely, on the fact that neutral hydrogen is present or not in the cell. 

\begin{figure*}%
    \includegraphics[width=0.33\textwidth]{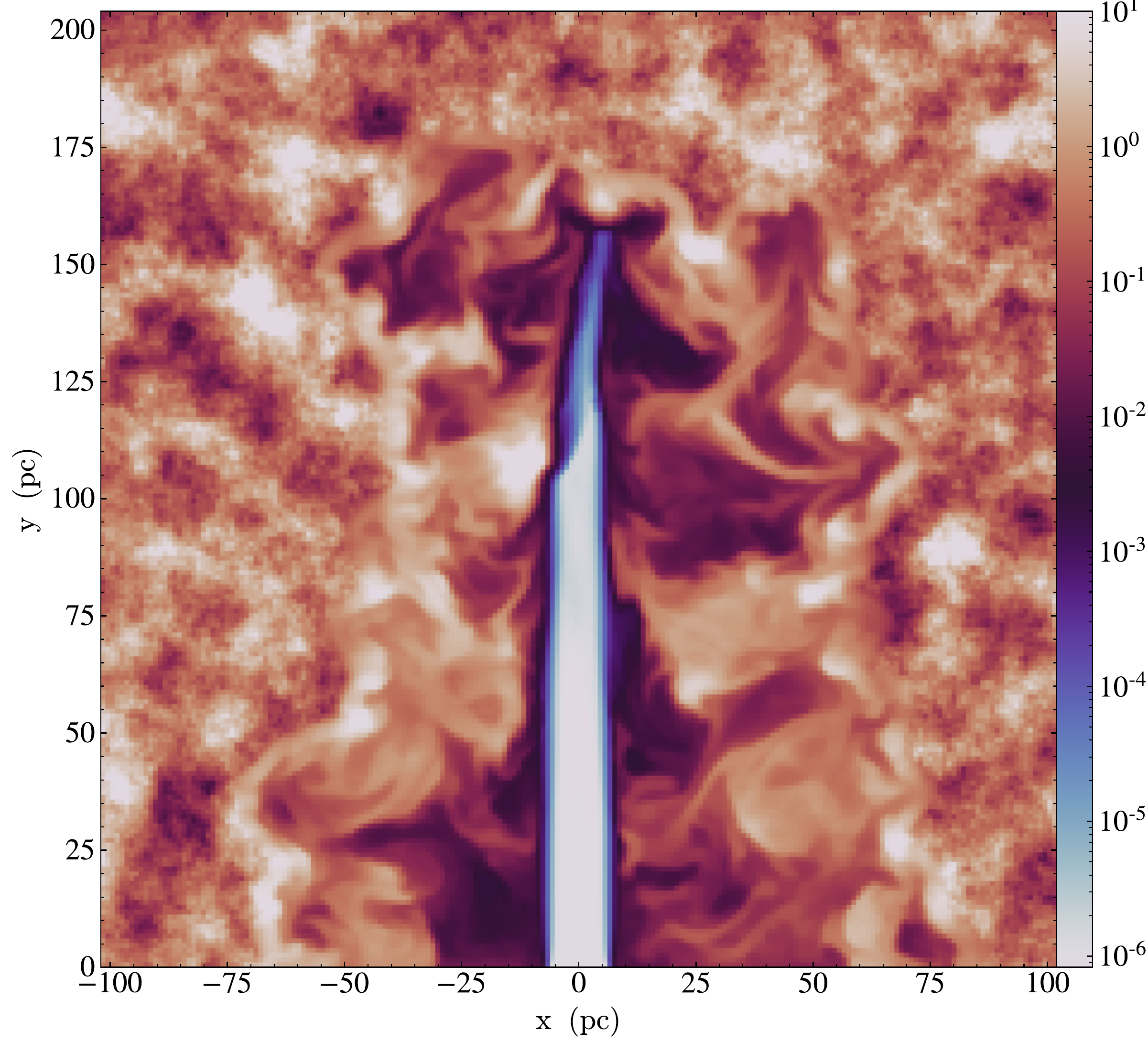}~
    \includegraphics[width=0.33\textwidth]{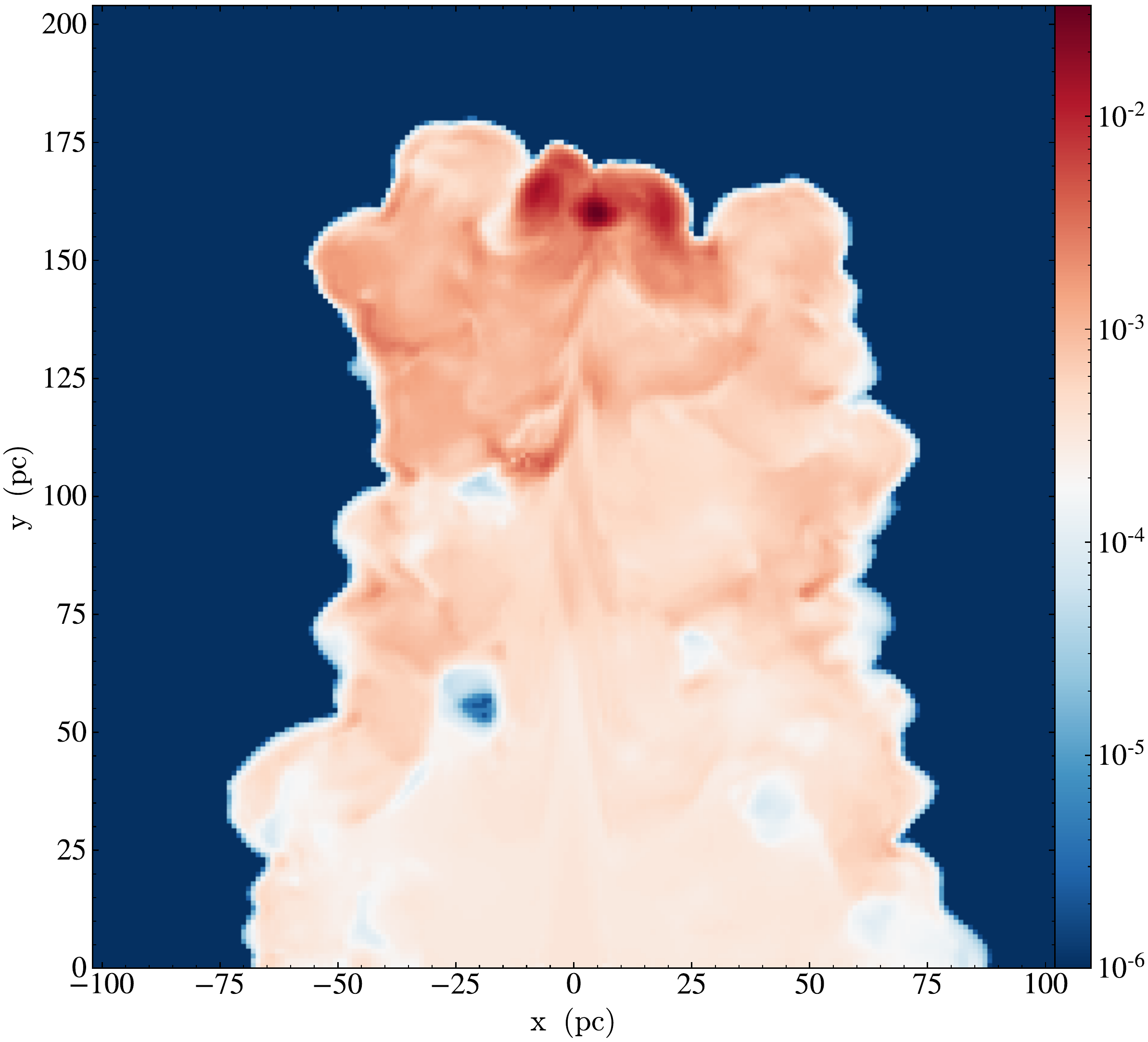}~
    \includegraphics[width=0.33\textwidth]{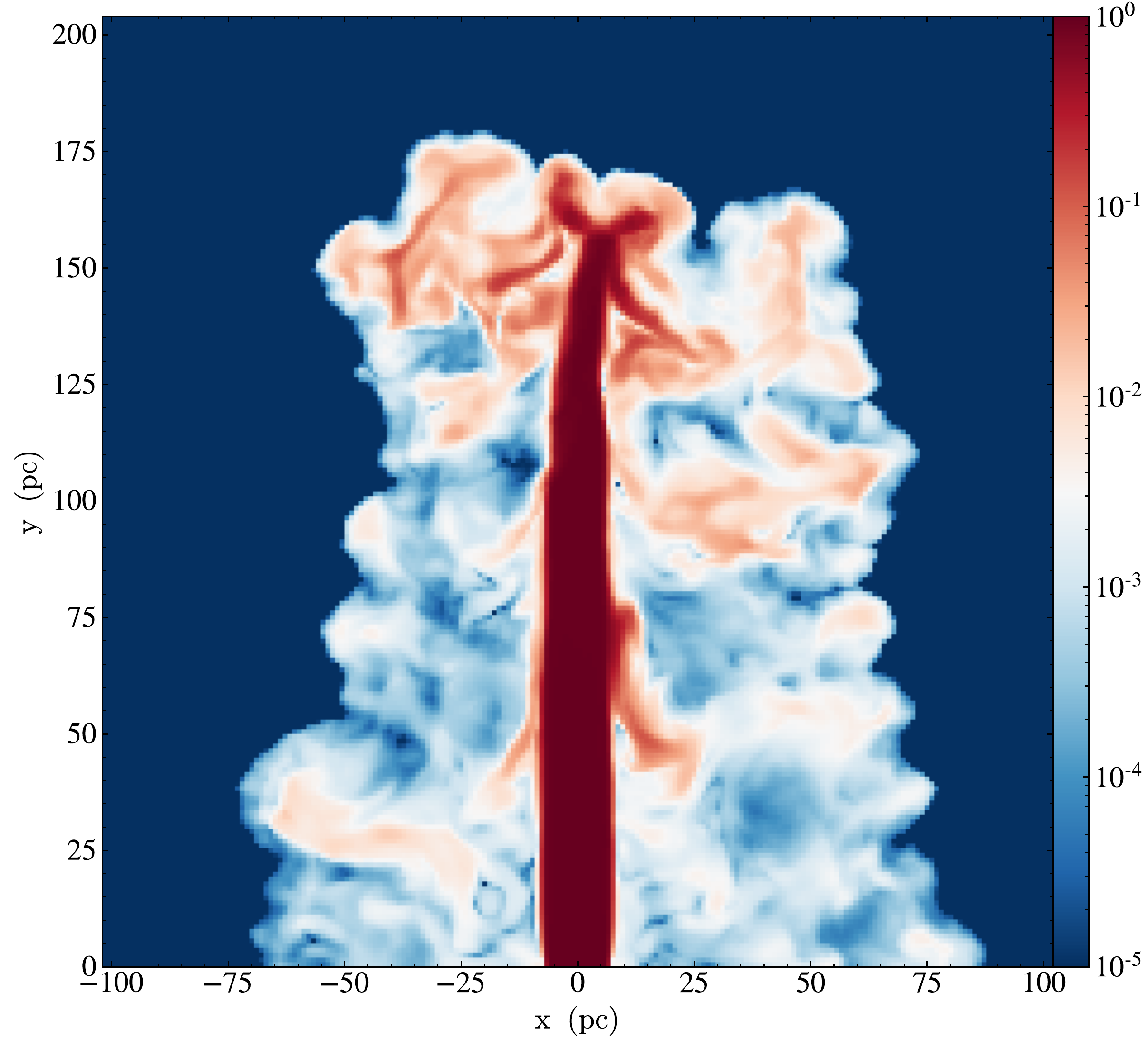}~

    \includegraphics[width=0.5\textwidth]{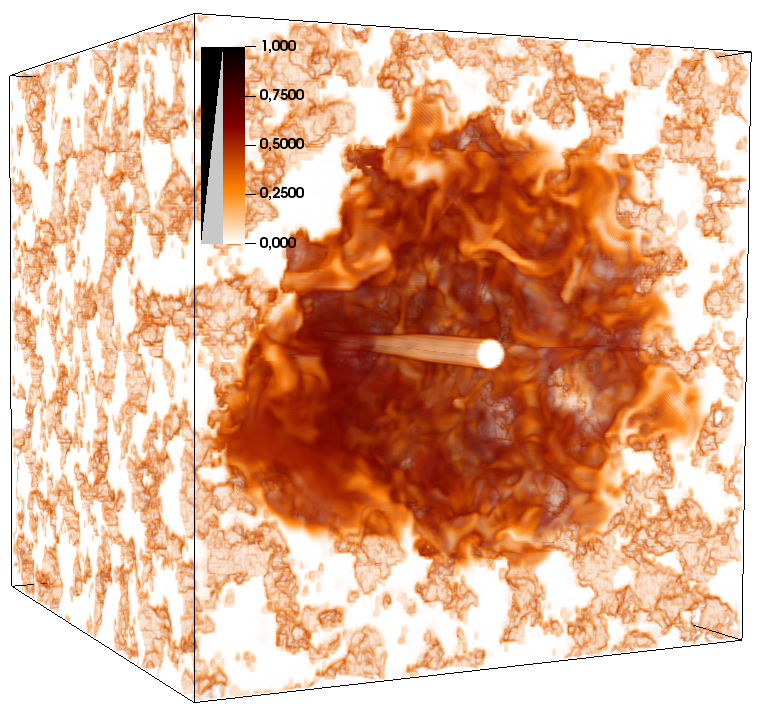}~
    \includegraphics[width=0.5\textwidth]{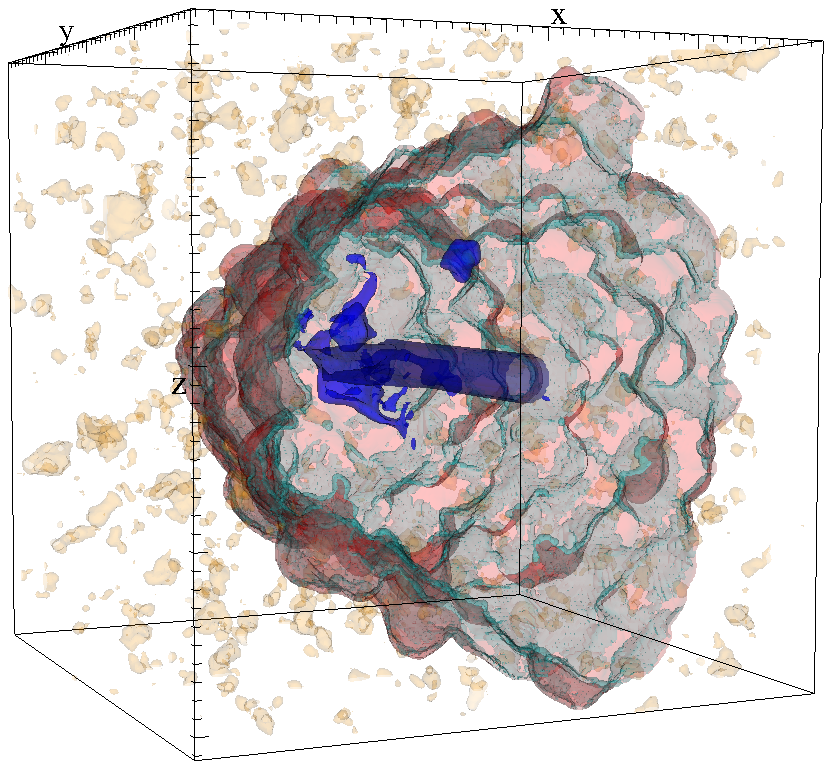}

    \caption{Top panels: Cuts of rest-mass density (left, in code units $\rho_a\,=\,10^7\,{\rm m_p/m^3}$), pressure (centre, in code units $\rho_a\,c^2\,=\,1.5\times10^{-3}\,{\rm Pa}$), and velocity field (right, in code units $c$) at $t\,=\,460\,{\rm yr}$ after injection. Bottom left: Tracer rendering showing the locations with originally dense, cloud, atomic gas (left). Bottom right: isosurfaces of pressure (red, $\simeq 10^{-10}\,{\rm Pa}$), leptonic number ($\rho_{e^-/e^+}/\rho = 10^{-3}$), atomic hydrogen density outside the shock (orange, $10^8\,{\rm m_p/m^3}$), and inside the shock (bluish, $10^5\,{\rm m_p/m^3}$).}
  \label{ion}
\end{figure*}

The top panels of Figure~\ref{ion} show cuts of rest-mass density (left, in code units $\rho_a\,=\,10^7\,{\rm m_p/m^3}$, with an upper limit to the colour scale set at $\rho_a\,=\,10^8\,{\rm m_p/m^3}$), pressure (centre, in code units $\rho_a\,c^2\,=\,1.5\times10^{-3}\,{\rm Pa}$, with a lower limit set at $\rho_a\,c^2\,=\,1.5\times10^{-8}\,{\rm Pa}$), and velocity field (right, in code units $c$, limited at $10^{-5}\,c$) at the last snapshot obtained for the simulation, which is still being run, at $t\,=\,460\,{\rm yr}$ after injection. These plots reveal a complex shock structure around the jet as forced by the strong inhomogeneity in density found in the ambient medium. The shocked region is also highly inhomogeneous in density, revealing the richness in the interaction between the jet and the ISM shocked gas. The high sound speed contributes to homogenize pressure within the shocked region with the exception of some knots of denser (ambient) gas, and the jet's hot-spot. Finally, the velocity field shows a wide range of values, from $10^{-4}$ to $0.1\,c$. Obviously, the smaller velocities correspond to the denser regions (see the comparison with the density panel). Although there are regions with values that can fall to several tens of km/s (blue color within the shocked area, $\sim 10^{-4}\,c$), the dominating blue-white transition implies typical velocities of the order of hundreds to a thousand km/s. These values are in agreement with the typical ones measured from line emission in jetted active galaxies \citep[see, e.g.,][]{2018A&ARv..26....4M,2021A&A...647A..63S}.

The bottom left panel in Fig.~\ref{ion} shows a limited box centred in the region occupied by the jet structure with a rendering of a tracer that indicates the regions where originally dense, cloud, atomic gas can be found. From the image, it is evident that this gas is fixed in clouds outside the shocked volume, but it is completely disrupted and mixed by the shock, and appears concentrated towards the shock region. Finally, the bottom right panel displays, for the same box, a collection of isosurfaces of pressure (red) to highlight the shock wave ($7\times10^{-7}\,\rho_a\,c^2 \,\simeq\, 10^{-10}\,{\rm Pa}$), leptonic number ($\rho_{e^-/e^+}/\rho$ at $10^{-3}$, as compared to $\simeq 5.4\times 10^{-4}$ for $e^-/p$ gas) to show the jet and still lepton dominated, unmixed, regions, atomic hydrogen density outside the shock (orange, for $10^8\,{\rm m_p/m^3}$) to show the cloud distribution, and inside the shock (bluish, for $10^5\,{\rm m_p/m^3}$) to show hints of atomic hydrogen within the shocked cavity. The low density (bluish) hydrogen density contours follow clearly the shock surface and thus reveal a very fast and complete ionization of the atomic gas. Therefore, we observe that the shocked gas is completely ionized, so despite our simulation recovers the observed velocities, as stated above, it shows that the shocks produced by powerful jets ionize atomic (and therefore molecular) gas very rapidly and this forbids line detection, unlike observations clearly reveal. The origin of this discrepancy probably lies in the fact that observed lines correspond to radio sources more evolved than our simulation at the current position (several kpc vs 200 pc) and that post-shock temperatures (typically $10^6-10^7\,{\rm K}$) are too high to allow for recombination of shock-ionized hydrogen within the simulated time. Nevertheless, we observe white spotted areas in the cuts (top left panel) which imply densities between $10^7$ and $10^8\,{\rm m_p/m^3}$ and this could translate in cooling times of $\sim 10^{3-4}-10^{4-5}\,{\rm yr}$ for gas between solar and 0 metalicity at the aforementioned temperatures \citep{2003adu..book.....D}. Although the simulations are limited to pairs, and atomic+ionized hydrogen in the composition of the gas, they show that the conditions for line emission with the observed velocities could be recovered once the jet has evolved for, at least, twice the simulated time so far. Therefore our simulations represent a promising first step in the understanding of ISM dynamics driven by jet injection in the host galaxies.  

\subsection{RMHD simulations}\label{sec:rmhd}

\subsubsection{Jet-stellar wind interactions}

Jets in X-ray binaries propagate through the binary environment before carving their way towards the ISM. In this region, specially in the case of massive binaries, jets travel through a dense, slow stellar wind that can produce a strong lateral impact on the jet flow, triggering internal shocks and, in the case of low power jets, even disruption. This was proposed as a plausible scenario for triggering gamma-ray radiation and thus explaining the emission physics of the very few gamma-ray emitting binaries observed in the Galaxy \citep[see e.g.,][and references therein]{rieger07,bosch09}. 

Following a number of papers that studied RHD simulations of jet-massive wind interactions \citep{2008A&A...482..917P,2010A&A...512L...4P,2012A&A...539A..57P}, we run a set of dedicated three-dimensional numerical simulations using our new code \textsc{lóstrego}, including the dynamical effects of magnetic fields \citep{2022A&A...661A.117L}. The results of these simulations showed that the latter could play a stabilizing role -and thus, to provide extra collimation within the binary scale- in the jet evolution when the flux of magnetic energy is low compared to the flux of kinetic energy, while a non-negligible magnetic energy flux makes the jet prone to the development of current-driven instabilities within the scale of the binary.

The dynamics of the jet model with a dynamically relevant field in \cite{2022A&A...661A.117L} is particularly interesting from several points of view. The jet evolution during the interaction with the stellar wind is significantly different with respect to a hydrodynamical jet with the same total power: whereas a powerful hydrodynamical jet evolves without losing collimation, the magnetised jet is prone to the development of current-driven instability pinching and kink modes triggered by overpressure and the toroidal field. Therefore, the jet shows a chain of recollimation shocks within the binary scale, followed by the development of a strong kink that spreads the jet momentum throughout a large area and decelerates its advance. The magnetic field structure is toroidal from injection to the development of the kink, while it becomes highly entangled near the head, filling the cavity. 

\begin{figure*}%
    \includegraphics[width=\linewidth]{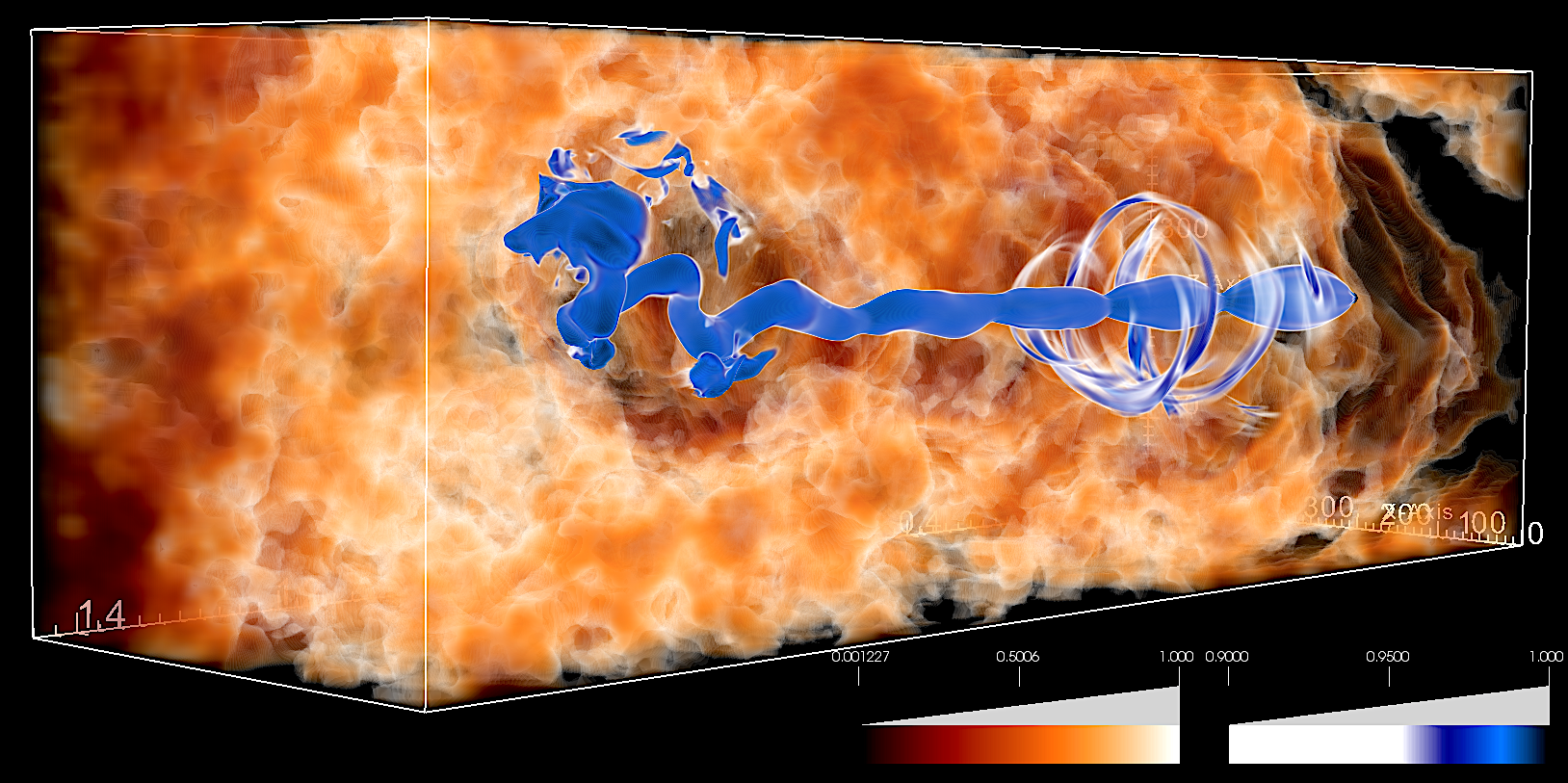}
    \includegraphics[width=\linewidth]{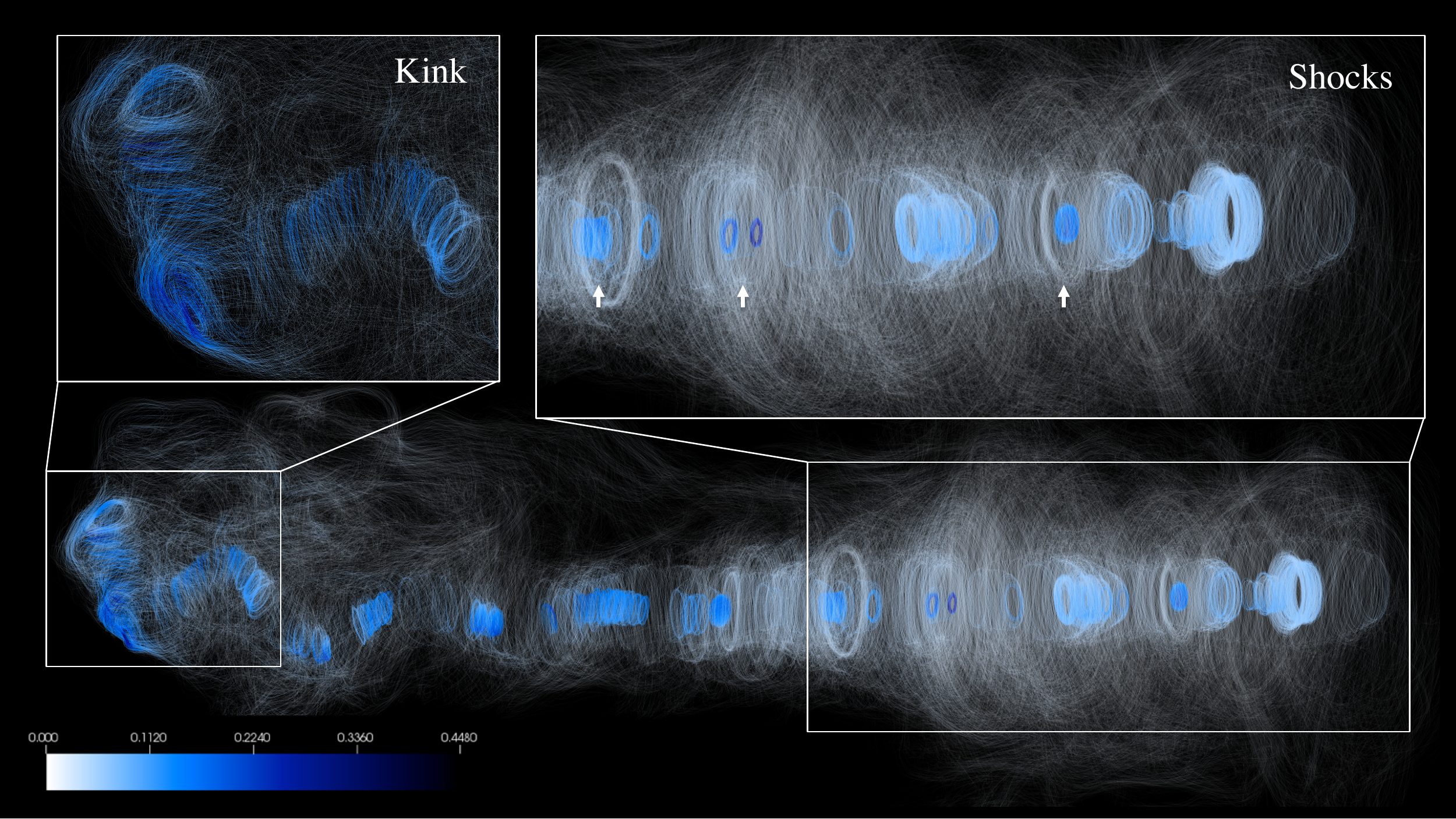}
    \caption{Top: 3D rendering visualization of the jet core ($f>0.9)$ propagating through the clumpy stellar wind. Bottom: Magnetic field lines integrated in the 3D volume, where the colour map represents the magnetic field vector module. The transparency of the lines is inversely proportional to the vector module to highlight the jet scheleton. Inset plots zoomed in the recollimation schock region (right) and the twisted trace triggered by the kink instability (left). }
  \label{kink}
\end{figure*}

In Figure~\ref{kink} we present the resulting configuration of the magnetic field in this simulation by focusing on the field lines. This plot shows relevant information about the field structure in the jet spine, which was hidden by the mixed plasma in the cocoon in our representation in \cite{2022A&A...661A.117L}. Since the magnetic field within the cocoon is low compared to that in the jet, we have applied a linear transparency to the field lines, weighted by the field vector module. In the mid-low part of the jet, the magnetic field preserves the toroidal structure of injection, which is reinforced at the recollimation shocks, as expected. The jet tracer 3D render (top panel), which is limited to $f>0.9$ to highlight the jet core, shows a pinched morphology and a set of annulus of jet plasma surrounding the core, which are deposited in the cavity due to the jet dynamics during the first stages of propagation. This morphology is very particular and could give rise to different emission patterns that deserve to be studied in detail, but this is out of the scope of this contribution. In the mid-upper part of the simulation, both the new representation of the field and the tracer function allow us to distinguish a well-resolved precessing morphology, triggered by the development of a kink instability. The inset plot of Fig.\ref{kink} zooms in the jet head, showing that the field is highly reinforced at the elbow of the twisted trace, which is also directly impacting the shocked environment formed by the stellar wind. 

Further analysis will be required to investigate the consequences of this precessing morphology on the jet dynamics (within and beyond the binary), on the one hand, and its implications in terms of non-thermal emission, on the other. This kinks could contribute to the observed signatures of precessing jets that have been resolved at radio wave-lengths for different X-ray binaries with jet-like structures. The most relevant example of jet precession is the well-known microquasar SS 433, but there are many other examples \citep[see e.g.,][]{mio01,massi12,miller-jones,luque20}. The origin of precession in microquasars is still debated for most of these sources. Although the classical theoretical models tend to explain precession by invoking relativistic effects of the inner disc \citep[i.e., Lense-thirring precession,][]{liska,motta} or the coupled effect of stellar winds and orbital motion \citep{maxbosch}, our simulation shows that current-driven instabilities can also trigger helical patterns during the jet-wind interaction, although the periods associated with this kind of precession should be accurately related. For example, current-driven instabilities have been claimed to explain the large-scale morphology of the jet structures observed in GRS 1758–258 \citep{luque20}.

This simulated jet thus represents an extraordinary candidate to produce high-energy radiation in microquasars, since energy dissipation may not only occur at the several reconfinement shocks where particles can be accelerated, but also a kink instability that can lead to kink-driven magnetic reconnection processes further out from the injection base \citep[see e.g.,][]{Bodo21,bodo22}. The characterization of the periodic behaviour of these processes (i.e., shocks and kink-driven precession) can be also relevant to interpret the rapid X-ray variability of some X-ray binaries/microquasars. Therefore, future work will include consideration of those dissipative processes in the acceleration of non-thermal particles and the triggering of high-energy radiative output. Although magnetic reconnection is not reproduced in ideal RMHD codes like ours, recipes as those presented in \cite{Bodo21} will allow us to characterize the formation of current-sheets and magnetic dissipation regions into our simulation. We expect that the combination of such analysis with numerical simulations as those presented in this contribution, will provide new insights into the physical processes that lead to non-thermal emission in microquasars jets within the scale of the binary.



\section{Conclusions}
In this paper, we have presented a number of scenarios that require the use of numerical codes to be studied due to their extreme nature, which makes this research prohibitive for current laboratory experimental capabilities. All these scenarios have in common the need for relativistic (magneto-)hydrodynamics, and they use similar methods to solve the system of equations that is used to model them. Different improvements have been made in the codes used in our group through the past years, including optimization and parallelization, modeling of the ambient media, etc. These improvements have allowed us to tackle more realistic simulations, which, although still far from reproducing the studied scenarios accurately, have brought relevant advances in our knowledge of the physics of relativistic outflows in Astrophysics and their impact on their environment. Future work should address in detail the role of magnetic fields in the evolution and impact of jets in the scenarios previously studied with RHD simulations while, at the same time, improving the physical processes that the codes take into account and that can be relevant for studying those scenarios. One of such improvements consists on including radiation dynamics, which could allow to compare the role of shocks and radiative output from the central engines (accretion disks) in shaping the heating mechanisms of the ISM and the IGM/WHIM.

Future improvements of codes will be related to link multi-scale processes and adding relevant physics to improve the simulations of astrophysical scenarios that are still extremely challenging for current supercomputing capabilities and far from realistic for laboratory experiments.



\section*{Acknowledgements}
Computer simulations have been carried out in the Servei d'Inform\`atica de la Universitat de Val\`encia (Tirant).  This work has been supported by the Spanish Ministry of Science through Grants PID2019-105510GB-C31/AEI/10.13039/501100011033, PID2019-107427GB-C33,  from the Generalitat Valenciana through grant PROMETEU/2019/071, it was also funded by the Deutsche Forschungsgemeinschaft (DFG, German Research Foundation) – project number 443220636, and it forms part of the Astrophysics and High Energy Physics programme and was supported by MCIN with funding from European Union NextGenerationEU (PRTR-C17.I1) and by Generalitat Valenciana.

\bibliographystyle{jpp}
\bibliography{numsims}

\end{document}